\documentstyle[12pt]{article}
\def\lsim{\lower.5ex\hbox{$\scriptstyle\buildrel < \over \sim $}}
\def\gsim{\lower.5ex\hbox{$\scriptstyle\buildrel > \over \sim $}}

\begin{document}

\begin{center}
{\Large\bf Creating chaos and the Life}\\[2mm]
           {\it Boris Chirikov} \\[2mm]
           Budker Institute of Nuclear Physics \\
           630090 Novosibirsk, Russia\\[2mm]
                      
\end{center}


\begin{center}
{\large Abstract}\\
\end{center}

 In this short report the first attempt of a new approach to the still
 mysterious phenomenon of the life, and its peak, the human being,
 is presented from the view point of the natural sciences, i.e.  of the
 physics in the broad sense of the word. This idea has come to my mind about
 10 years ago when doing a completely different problem I suddenly have noticed
 to my surprise (see [1], p.20) a wonderful relation between a very complicated
 human (physical) conception {\it creation} and the relatively simple
 mathematical theorem due to Alekseev - Brudno  (see for instance [2]) in an
 almost unknown for physicists field of the so-called {\it symbolic dynamics}
 and {\it algorithmic chaos}, which one I have immediately christen
 {\it the creating chaos}.
 
 Unfortunately, realizing very well an incredible complexity of all the life
 and especially the present human being, homo sapiens, as well as my own 
 complete ignorance in this field, I had to put aside any farther studies
 of my "creating chaos" for indefinite time.
 
 However, to my surprise, it gradually turned out that not only me, a curious
 physicist, beginner in this field, but honourable psychologists as well can
 neither understand, nor explain what is after all the principal distinction
 between the human being and the ape (as a generalized representative of the
 common highest primates) in spite it is really striking. This problem is
 well known to experts, yet it is not only remains unsolved but, to the
 contrary, is beclouding by somewhat misty hints on a possible wisdom of the
 ape comparable with human talent (see, for example [3]) !

 Here I had just wised up, that the time of my {\it creating chaos} has come.
 Certainly, some conception of chaos, probability and statistical laws have
 been known since long ago, as the random mutations of genes, for example.
 Yet, that was not the chaos which could explain the laws of the life, so
 different from anything else (see for example [4,5]).

\newpage

 {\bf 1. Creating chaos in classical dynamics.}
 A relatively simple and detailed physical description of the symbolic dynamics
 is presented in [1], including the most interesting, in the problem 
 under consideration, mechanism of {\it random symbolic trajectories}, or
 {\it the algorithmic chaos}. For this problem the most important peculiarity
 of the {\it algorithmic chaos} is in the statistical properties (fluctuations)
 of {\it individual trajectories} of a single particular system instead of
 averaging over an arbitrary ensemble in which the largest and most important
 fluctuations are hopelessly lost. To avoid this one can make use instead 
 of the "true" continuous trajectories the so-called {\it symbolic 
 trajectories} which are presented by the projections of the former on a 
 certain discrete lattice in the system phase space, and moreover, in the
 particular, also discrete, instants of time. A finite time step of the 
 trajectory allows to introduce a specific conception of the {\it complexity
 of symbolic trajectory} (after Kolmogorov) which, under the certain 
 conditions, serves as the ideal {\it algorithmic information} (after Chaitin
 [6]), the main statistical characteristic of the symbolic dynamics, at least
 for physicists.
 
 Most interesting and surprising here is that under the certain conditions
 the dynamics of such system becomes extremely chaotic in the sense of absence
 of any correlations in time. In other words, almost any individual symbolic
 (discrete) trajectory of such a system becomes absolutely unpredictable,
 i.e. it is impossible in principle to calculate almost any of its (discrete)
 values, even if all the previous as well as succeeding discrete values are 
 known exactly ! Such is the effect of incommensurability of the dynamical
 scales for the continuous trajectory and its discrete symbolic projection.

 In a reduced form {\it the symbolic dynamics} is described by a rather simple
 equation (see [1], Eq.(2.19)), which can be represented as
 $$
   \lim_{|t|\to\infty}\,{I(t)\over |t|}\,=\,h\,=\,\Sigma\Lambda_+ 
 $$
 where $\Lambda_+$ are the positive Lyapunov exponents, characterizing the
 local instability of the limiting continuous trajectory and their sum $h$
 stands for the metric Kolmogorov - Sinai entropy. The latter characterizes
 the most important in the problem parameter, the complexity of a single
 symbolic trajectory, or {\it the density of algorithmic information} about
 trajectory on the time interval $|t|>0$.

 This is still some simplified mathematics. The physics appears in my 
 hypothesis [1], that the algorithmic information is just the qualitative
 characteristic of creation. The latter is always present in one form or 
 another for any model parameters. However, a continuous unrestricted creation
 requires extremely strong, exponential instability of the system:
 $$
   h\,=\,\Sigma\Lambda_+ > 0
 $$
 On the contrary, any regular sorting out any parameters of the system,
 including any external interactions is going to certainly suppress any 
 creation after all.
 
 Moreover, for application of the mathematical method of symbolic 
 dynamics in physics the complete absence of correlations in algorithmic 
 information is required, at least in the limit $|t|\to\infty$.
 
 This is just the end the discrete symbolic trajectories are used which,
 in the chaotic regime $h>0$, produce the complete {\it indeterminism} of
 individual trajectories, i.e. rule out any systematic repetition of the
 algorithmic information, which would be a contradiction with our intuition
 of unrestricted creation.
 
 In spite of unusual terminology for physicists, the physical meaning of
 symbolic dynamics is actually very simple. This is just a specific observation
 (measurement) of system motion by a human being, human itself or its special
 device which results only human can use for its own purposes,
 particularly for the science, discovered and developing by the human itself.


 {\bf 2. Creating chaos in quantum dynamics.}
 Until now only classical models have been mentioned for simplicity. However,
 by the first look, the biological systems seem to be essentially quantum
 ones. If so, is there any physical meaning in the study of classical models ?
 As a hypothesis it is not completely excluded since in both cases the main
 point remains the same, the external with respect to a living system 
 measurement by the macroscopic observer, the human being. Importantly,
 that in this case it is not at all the object to study. On the contrary, it
 plays a purely ancillary role, absolutely necessary but still an ancillary 
 one. Without direct continuous participation of the human it would be
 impossible to develop any science, any natural science (= physics, see
 beginning of this report), even though the human itself  does not directly
 appear. Otherwise, one can say that the human being plays in science a double
 role:

 (1) as a certain particular physical system, the extremely specific and
 complicated "mote" in our Universe, the mote of paramount importance for us
 as human beings but not as physicists since this "dust" does not at all
 influence any physics of the World in spite of the so-called {\it anthropic
 principle} in the physics itself (see for example [7]) which is essentially
 correct but rather misleading in formulation, and

 (2) as unavoidable "construction" of the whole physics irrespectively of the
 existence of human being itself, particularly before its appearance (!)
 during the evolution of the life (see for example [1] and Section 3 below).
 
 It is just this second "hidden" role of the human being, particularly the
 future human, which still brings some physicists to an "evident" confusion,
 especially in quantum mechanics (see for example a detailed discussion in [8]).
 In any case, a serious question arises as to an ambiguous human construction
 of science, that is well known in practice of the science creation (see for
 example [9]). Moreover, if the creation itself turns out to be chaotic indeed,
 the "construction of science" becomes completely indefinite ?? This important
 question will be also considered in Section 3. So far, one can only remark
 that in mathematical language there is no serious cause to expect any
 considerable discrepancy between classical and quantum model which corresponds
 to the same method of symbolic dynamics in both cases. Perhaps, this is just
 the reason why physicists don't like this mathematical invention, which may
 look for them "suspicious". The point is that until now the characteristic
 quantum indeterminism, related just to the measurement, is regarded as an
 exclusive peculiarity of the quantum mechanics only.
 
 And suddenly, the unexpected discovery: a complete classical model of quantum
 measurement ! The discovery has been made recently [10,11] by famous soviet
 (american since the end of 80th, expelled from USSR [12]) physicist Yuri Orlov
.During many years, in unbelievable soviet conditions he was theoretically
 studying, in the frame of quantum mechanics, the physics of human psychics, i.e.
 the physical processes, designating the behavior of human both a single one
 as well as in a collective. One of the main problems of his studies was the
 so-called {\it human freedom of will}, its mechanism and ability. Already half
 a century ago Schr\"odinger was fantasizing about this still unsolved mystery of
 the life in his remarkable book on the physics of a living cell [5].
 This is how he determined his curiosity to this problem in the closing chapter
 of his book

 {\it On determinism and the freedom of will}

 "In reward for the work in presentation of the purely scientific side of our
 problem...I beg now for permission me to express my own, inevitably subjective
 view at the philosophical value of the question...I beg the reader to give up,
 for a time...some particular opinions, and to consider an essential point...,
 what could be the contribution of biologist, trying to prove in one strike
 both the existence of god and immortality of psyche".
 
 Orlov's problem was and remains an incomparably more complicated: to prove
 within the physics, not a philosophical fantasy, a possibility of 
 indeterminism and the freedom of will in human being and only the human being.
 And he has got the interest not simply to a problem of the fundamental science
 but also as an application to the politics (!): what is the freedom of will ??
 the freedom of choice or "the realized necessity" ? - a standard slogan of the
 soviet philosophy, attributed to Lenin even though it has been used and much
 earlier in the world philosophy but certainly not in such a categorical form
 ("old wives' tale on the freedom of will").
 
 Generally, it is very important to clearly separate the truth of science
 (physics) which is "ratified" by the experiment only and remains forever
 from various "surmises" of the so-called "humanitarian sciences" 
 (see next Section), very diverse in content and strongly fluctuating in time.
 True, all these fantasies may turn out to be extremely useful prompts for
 the development of physics, yet they never are deciding because of their
 undetermined nature. A fresh example is the second, auxiliary, role of the
 human in symbolic dynamics where the human being represents not the content of
 the physical law but its symbolic construction only (a pun in symbolic
 dynamics, see Section 1).

 Curiously, a prompt here could be (?) the final (critical) Kant philosophy 
 on the
 "thing-in-itself", unobservable by the human, that almost literally corresponds
 to the present mathematics and physics of symbolic dynamics under conditions
 of chaos and indeterminism (see for example [1], p.20). Thus, philosopher Kant
 had got ahead of mathematician Hadamard by more than 100 years and of 
 contemporary
 physicists, in the best case for them, by 200 years !
 Remarkably, that even 
 though Kant's aim was some (any ?) justification of demand on human for the
 moral (which one ? any ??), he intuitively understood the necessity of
 "freedom-of-will/indeterminism" combination for the human moral 
 responsibility.
 
 On the other hand, Orlov was the first who really has {\it proved}, in the
 frame of present physics, the principal possibility of such a combination
 [10,11]. Strikingly, he did this fully independently of the well known 
 mathematical method of symbolic dynamics ! Naturally, he was using a rather
 different terminology, including, particularly, and some strange expressions
 like, for example "indeterminism without chaos" [11]. Only after fairly
 prolonged discussions during the Novosibirsk conference [11] we both
 (I believe !) have managed to understand the very important correspondence
 between Orlov's physical theory and the mathematical method of symbolic 
 dynamics .
 
 Mathematically, both models are equivalent, that justifies using the 
 classical model of symbolic dynamics, at least at the initial stage of
 studying the creating chaos of the life. Such a mechanism, classical and
 quantum, is really possible. However, is the homo sapiens and its brain 
 actually constructed like this, still remains to be understood.
 
 To avoid confusions one should note that strange at the first glance 
 equivalence of classical and quantum models of the system in question
 is explained by a peculiar representation of both models which dynamics
 is not a "true" one but only their macroscopic observation (measurement).
 In the classical model (classical limit of quantum mechanics) the full
 dynamics of a continuous trajectory can be introduced independent of the
 measurement normalization, i.e. as invariant with respect to the observer.
 However, in quantum mechanics this question still reminds open in my
 opinion. In any event, the noninvariant measurement is certainly related 
 to a very specific noncoherent state of the quantum system during
 measurement.
 
 A living system open to unrestricted interaction with external sources could
 never be transformed into a coherent state of a partly alive "Schr\"odinger cat" !
 

 {\bf 3. Invariance of the Laws of Physics ?}
 Would it be possible to develop the invariant Law of Physics in spite of arbitrary
 actions of the human being as an auxiliary element of the science construction
 (see Section 1) ? In a due time, quite a long ago, this question was put and
 resolved by Einstein already. Today, his philosophy (logic) of Science could
 be presented something like this: "The human Science is its model of the real
 World".
 
 This Einstein's credo, rather well known in science, is the best resolution,
 in my opinion, of the closed circle of the "mutual affect" between a 
 negligible "mote" in the Universe, a human being, and its Science which Laws
 control the whole Universe including the human itself. In other words, such a
 philosophy naturally separates the natural sciences in noninvariant human 
 model and invariant, with respect to the model, the Law of the World. 
 Thus I myself, as well as many others, do include in the conception of Science
 the Physics only (in the broad sense of the word) which contains all the other
 Natural Sciences like chemistry and biology but not humanitarian ones
 ("unnatural sciences" in a witty remark by S.P. Kapitsa), such as philosophy,
 philology, economy, sociology, literature, history, jus, informatics,
 diplomacy, politics, in short, all that directly or indirectly depend on
 the so-called human freedom of will (see Section 5) and, hence, cannot be
 invariant with respect to human models. In other words, "Supreme verdict of
 Experiment" is completely irrelevant to humanitarian "knowledge".
 Instead, the human can only classify it or as its own "conditional agreement",
 for example, philology and even informatics (?), or as emotional favor
 ("culture"), or, at last, as "persistent" attempts to guess the "genuine"
 interests and wishes of the humankind, which moreover are always changing.
 Curiously, such a separation of some humanitarian disciplinaries has entered
 even certain languages, the English, for example (see any Webster's), yet not
 the Russian, in any case not the soviet Russian ! Notice a clear sign of the
 conditional character of humanitarian "laws" or, at least, of a totally
 special form for some of them. Particularly, a "fresh" defence of quantum
 idealism [13], as a part of philosophy, has nothing to do with any definite
 science (quantum physics). Notice, that one should not to confuse it with
 the general human construction of science as discussed above. This is well
 known, of course (Section 2), but right now there is a fresh argument also:
 the equivalence of observed (measured) classical and quantum indeterminism
 (chaos) as discovered by Orlov [10,11] (Sections 2 and 6).
 
 Coming back to the problem of invariance of Physics, one should also keep
 in mind the importance to become free from seeming influence of the human on
 the whole Universe. This still leads to, a hidden though, revival of ancient
 mystics while it could be simply an artefact of the structure
 of human science (see [7] and Section 2). Unfortunately, such Einstein's
 logic still has not yet received any development because of a formal
 contradiction with "basic" philosophy of the quantum mechanics which 
 "forbids" using nonobservable variables in physics. It is not excluded that
 now, after the Orlov discovery, the relation of physicists to the Einstein
 logic would also change. In any case, there seem to be here no serious 
 contradictions so far. 
 

 {\bf 4. Evolution of Life prior to the human but according to its Science ! }
 On this initial stage of the life development the creating chaos remains
 the external one with respect to biological objects. Generally, it is simply
 heat fluctuations which produce some random changes, particularly making the
 molecules of environment more complicated. The critical point of the birth
 of life may be accepted as the appearance of the so-called {\it Darwin's
 triad} (see for example [5]): heritage (providing a long stability of life),
 variability (providing a slow evolution of life), and selection (natural,
 providing a fast removing of noncompatible competing species).
 
 In the frame of my hypothesis for the physics of life, the evolution of life
 seems to be the most interesting as it looks so surprising comparable to other
 known physical processes that it still is considered by someone as incredible
 mystics. Roughly, there exist, at least, two
 essentially different processes of the life evolution. One of them, more
 simple and well studied, could be termed a local evolution. It is the
 evolution, which responds to restricted variations of the life in a regular
 way. In this case the variability of Darwin's triad simply means a
 sufficiently quick many-variant adaptation to the new conditions (see for
 example [14]).
 
 Another evolution, practically unstudied yet at all, which goes on 
 simultaneously,
 is much more interesting. Generally, this global evolution does not depend
 on the life conditions but it is just the evolution which determines the
 contemporary "miracle" of the life with extremely complicated structure up to
 homo sapiens itself.

 A detailed historical review of both evolution regimes, Darwin's selectionism
 and neocatastrophism with sudden births of species, is presented in [15],
 including the many-year "opposition" of these two scientific directions, which
 is still not yet resolved. Nevertheless, the author [15] discerns a perspective
 of a new synthesis of both directions using most recent and most various
 achievements of the natural sciences, all well known but one which is still
 completely missed. This is the creating chaos, based on also widely ignorant
 symbolic dynamics beside some mathematicians.
 
 Meanwhile, the creating chaos (indeterminism) is just the process which allows
 a natural explanation of sudden births of unrestricted species produced by
 an external symbolic chaos (Sections 1 and 2). If it is really in agreement
 with enormous empirical data collected during the epoch of "opposition", this
 could be considered as the first conformation of the creating-chaos 
 hypothesis, which is not only possible in physics but also does work, as a 
 matter of fact, in the Life, at least prior to the human. Essentially, that
 conclusion cardinally depends on the existence of classical model for the
 quantum measurement discovered by Orlov (Section 2). Very important by itself,
 this discovery allows for a direct relation of a seemingly pure quantum
 indeterministic measurement with the classical conception of creation which,
 in my opinion, plays the central role in the whole problem of the Life.
 
 Interestingly, the cardinal role of discrete quantum spectrum in biology,
 producing "leap-shape" mutations [5], is essentially the same as a symbolic 
 discrete trajectory of the classical model, i.e. in both cases the observable
 part of dynamics only. Actually, what is really important is not the discrete
 spectrum in quantum model or a discrete classical symbolic trajectory but
 the possibility of unrestricted creating  Nature and,
 hence, the evolution of Life. A quantitative estimate of the evolution
 parameters is determined by simple classical equations in Section 1.
 

 {\bf 5. Human kingdom under chaos control ?}
 Now, let's turn to the top of the Nature creation, the human being, 
 contemporary human, homo sapiens, who is cardinally different from any other
 biological object, including the most close to the human the highest primates,
 for example the ape. Nowadays, one doesn't need to be an expert for clear
 seeing the human "might" uncomparable with anything else. What is the origin
 of such a "might" and where it is going to bring us in a near future ? 
 This is one
 of the most fundamental problem of biology or, more precisely, of 
 contemporary  psychology. This problem is well known to the experts, yet
 nobody even tries to solve it. In my impression, the main obstacle in
 this and many other scientific problems of the Life is a common ignorance
 of a relatively new phenomenon of complete indeterminism (Orlov [10,11]
 or creating chaos [1] or symbolic dynamics (Sections 1 and 2)) while the
 "ordinary" chaos doesn't help at all in this case [4,5]. Meanwhile, not only
 biologists but even physicists (!) pay no attention to this new part of their
 science.
 
 In any way, the human is cardinally different from all the other Life by
 its {\it freedom of will} which we directly feel and know simply from our
 own experience and our personal unique $\cal I$. Actually, after creating 
 the human, the Nature
 has deprived it of the famous Darwin's triad  based on  the dynamics of
 biological species and, hence, of the biological evolution as well. With the
 freedom of will  the human has become its own lord, and seems to be already
 ready to make the same with the famous Vernadsky noosphere.
 
 Unlike Darwin's triad, the human freedom of will, if it really exists according
 to my hypothesis, is determined by an internal rather then external creating
 chaos. Roughly, the mechanism of human creation is related to the structure
 of its own brain. This structure is rather similar for all the highest primates
 beside the two main differences.
 
 (1) The human language for the wide external information exchange among many
 people and for the internal thinking of a single person. Somebody considers
 this as the cardinal one. In essence, I cannot say anything here except
 that in the frame of my hypothesis the second one seems to me more
 important.
 
 (2) Only the human brain has asymmetry between the both semi-spheres, a very
 important one, not geometrical but deeply functional.
 
 This asymmetry is so serious that in contemporary psychology the following
 strange paradox has immediately appeared and is still continuing for quite 
 a lot of time.
 
 On the one hand - the well studied left brain with a relatively simple
 structure and a regular operation, which however does not single out the
 human within the animals (beside the language), but on the other - a "zest"
 (the priceless pearl)
 of homo sapiens, the right brain, which is so complicated that the experts
 have prefered to forget this to remain "experts" !
 
 The result ? A science of animals, including the human as well, and a pure
 empirics for homo sapiens to control the human by human ??? Certainly,
 I somewhat overcolour in the sense that the creation by the human itself as
 a product of its right brain and the basis of its "might" are qualitatively
 known and accepted, at least by some experts. However, farther the science
 has stopped ! The reason is the same - a new unusual chaos, the complete
 indeterminism is required which remains unknown   to physicists at all
 (why a new one if there is well developed old ?). Moreover, it looks that
 even mathematicians don't understand well their new chaos, at least with
 respect to the applications in physics.
 
 Here are a few words by Poincar\'e - philosopher from the collection of his
 publications and speeches on the philosophy of science [16] (p.515).
 
 "Even though I have to finish my considerations soon, I cannot pass over one
 important point in silence. The science is deterministic, it is such a priori,
 it does postulate determinism since it could not exist other way. It is such
 a posteriori as well; if it has postulated this from the very beginning as a
 necessary condition of its own existence, then later it rigorously proves
 this by its own existence, and each of its victory is the victory of
 determinism.. This is the question which has been studied without any success
 in many centuries, and I cannot even to present it in a few minutes I have".
 
 Poincar\'e - philosopher just have missed a new mathematical construction
 developed 10 years before (in 1898) by another mathematician and philosopher
 Hadamard 
 which {\it under particular conditions} excludes determinism completely but
 not at average only as the usual statistical laws do, accepted by both
 Poincar\'e - philosopher and Poincar\'e - physicist. Apparently, Hadamard
 himself also missed this phenomenon, at least in applications in physics.
 Even contemporary mathematicians, to say nothing about the physicists, don't
 seem to be much curious to a more deep picture of this strange symbolic
 dynamics. For example, in a mathematical paper [2] one can read:
 .."quasi-random" behavior ..within the framework of mathematical determinism,
 i.e. the uniqueness of the solution of Cauchy's problem. .. An apparent loss
 of determinism .. caused not by random intervention, but rather by our
 assumption of impossibility of a precise determination of the position of
 a phase point". 
 
 I would like to stress again that the scientific foundation of the human
 Might is the {\it Freedom of human Will}, the real freedom, the freedom of
 selection as Orlov conjectured for the human (see Section 2), and as he
 proved for a simple example of quantum system [10,11]. It is another problem
 if the human brain does correspond to this example ? In any case, my impression
 is that to the present time there are already no serious reasons to expect any
 essential differences between the physics of biological objects and the other
 structures in the same domain of the parameters. The evolution of life,
 including the big leap to the human, is certainly not the Big Bang of the
 Universe or its inflation. All that is already behind ! 
 
 The key point is the freedom - of - will reality for the human, its moving
 force, which for Schr\"odinger was a subjective philosophy only (Section 2),
 the fantasy on the existence of god and immortality of psyche, never confirmed.
 Unlike this, in physics the confirmation is possible and necessary, via
 experiment (Section 3). And if this will be confirmed indeed, one more, perhaps
 the most dramatic one, problem is going to start up in a near future, the
 so-called {\it problem of sustainable development of mankind} (see for example
 [17]). The point is that (complete) freedom of will for the human means the
 (complete) absence of any systematic recurrences to the acts of this freedom
 (Section 1), hence a random, completely unpredictable behavior of the human 
 with
 all its "might" over a sufficiently long time $|t|\to\infty$. 
 
 A new paradox again ! On the one hand, the human is "lord and king", and can
 do anything whatever it does "wish", but on the other, it does this completely
 at random, "not wising what creating".
 
 The result ? A random wandering of "King and its warriors", that eventually
 will bring all the humankind to a big catastrophe. This danger is well
 understood by even a rather restricted science of Life, and apparently by
 the world public trying last time some active, I would say agonized, attempts,
 including the United Nations, to find the solution. Preliminary, too
 preliminary, the expected solution has been termed "sustainable development of
 mankind" (see for example [17]). Here, it means the realization, or rather
 an attempt of that, of a certain agreed action of the whole mankind for
 realization of a certain selected end.
 
 However, under condition of the freedom of will such a behavior of people
 is simply impossible as it is unpredictable by definition.
 
 In principle, the catastrophe could be push away for a finite time, if one
 would manage (sufficiently fast !) to understand its nature and mechanism for
 some corrections of the chaotic acts of the will. However, also in this case
 the serious difficulties arise as well concerning coordination of the will
 even of very, very but various, various as well.
 
 Does it mean, that all the actions of a human, even the most "genial" one are
 a simple chaos, including the very unusual, symbolic ? 
 
 Yes ! If this is the action of the human itself, its main role in life
 (Section 2). Then, it is not simply chaos but a creating chaos, the inimitable
 creation of any {\it particular human} with its own $\cal I$ ! What is here
 about "the sustainable development of the whole mankind" ?
 
 Never ! If those are just symbols of the human only, in the discovered by it
 and developing Science checked and "affirmed" by experiment (Sections 2 and 3).
 Then, the laws of Science themselves can be or not to be chaotic ones,
 depending on various conditions. Yet, these are the fundamental laws of Physics
 (the natural laws, Section 3), including the specific laws of Life and the
 human itself. The last laws, unlike the first are not universal at all but,
 to the contrary, in the framework of the accepted hypothesis, do represent
 the creating chaos, or the complete indeterminism. Generally, it is true
 in the limit $|t|\to\infty$ only, while over a finite time $|t| < \infty$
 some control of the creating chaos is possible, and even its regularization
 via the left brain, for example (see above and Section 1).
 
 The trouble of contemporary psychology is just in that the latter, relatively
 simple and very wide-spread, actually the only using method, turns out to be
 the least efficient (see above). Moreover, as a matter of fact, it reduces
 the human to the "ape" ! 
 
 To the contrary, the source of the human "might" is in the right brain which
 should be used for the correction of asymptotically unpredictable human
 behavior. However, to do this the human needs to understand how does work this
 most complicated and still mysterious part of its own brain.
 
 And what is not less important for both the researchers of the science of life 
 and its practitioners, is to get rid of a {\it wrong} impression that the
 chaos control on a finite time could be extended somehow to any 
 $|t|\to\infty$. This certainly disagrees with the symbolic dynamics 
 (Section 1), which conserves in the limit the maximal chaos, or the complete
 indeterminism, only, i.e. the unique creation of the Nature or that of the
 human itself. 
 
 In my view, this is the main lesson for the human from its own studies of the
 creating chaos, no matter do physicists like that or not.

\vspace{3mm} 

 {\bf Instead of conclusion}
 
\vspace{2mm} 

 My crucial question on the still mysterious Life:
 

 {\bf what differs human from the "ape"}.
 
\vspace{2mm} 

 My answer: 
 

 {\bf the freedom of will and its complete indeterminism}.
 
\vspace{2mm} 

 That is the source of unrestricted "might" of the human leading to inevitable
 catastrophe on our little Earth. In spite of all increasing self-delusion, the
 {\it homo sapiens} has hardly enough time to rescue itself from it own.
 My only hope is my own crucial mistake !? But where it is if any ???


\vspace {2mm}

\begin{enumerate}\item{} B.V. Chirikov, Natural Laws and Human Prediction, Proc.Intern.Conf.
           "Law and Prediction in the Light of Chaos Research"
           (Salzburg, July 1994),
           Eds. Paul Weingartner and Gerhard Schurz, Springer, 1996, p. 10.
\item{} V.M. Alekseev and M.V. Yakobson, Symbolic Dynamics and Hyperbolic
             Dynamic Systems, Phys.Repts. {\bf 75 }, 287 (1981).
\item{} P.V. Simonov, Creating brain: Neurobiological foundations of
             creation, Moscow, NAUKA, 1993 (see p.9 ???), in Russian.
\item{} M. Eigen, Naturwiss. {\bf 58}, 465 (1971);
           Molecular selforganization and early stages of evolution,
           Usp.Fiz.Nauk {\bf 109}, 545 (1973).
\item{} Erwin Schr\"odinger, What is life ? The physical aspect of the living
           cell, 1945.
\item{} G. Chaitin, Algorithmic Information Theory, in: {\it Information,
           Randomness and Incompleteness}, World Scientific, 1987, p.38.
\item{} B. Carr \& M. Rees, The anthropic principle and the structure
           of the physical world, Nature {\bf 278}, 605 (1979);
        C. Hogan, Why the universe is just so, Rev.Mod.Phys. {\bf 72}, 1149
           (2000).
\item{} M. Gell-Mann and J. Hartle, {\it Quantum Mechanics in the Light
            of Quantum Cosmology}, Proc. 3rd Int. Symposium on the Foundations
            of Quantum Mechanics in the Light of New Technology, Tokyo, 1989.
\item{} A.B. Migdal, The search for truth (some notes on the Science creation),
             Moscow, ZNANIE, 1978, in Russian.
\item{} Yu.F. Orlov, Origin of Quantum Indeterminism and Irreversibility of
                Measurements, Phys. Rev. Lett. {\bf 82}, 243 (1999).
\item{} Yu.F. Orlov, Indeterminism without Chaos: Classical Systems with
              Quantum Properties, Intern. Conf. "Dynamical Chaos in Classical
              and Quantum Physics", Novosibirsk, Budker INP, 4-9 August 2003.
\item{} Yu.F. Orlov, Dangerous ideas: memoirs on the Russian life, Arguments
              and Facts, Moscow, 1992, in Russian.
\item{} M.A. Popov, In defence of quantum idealism,
             Usp.Fiz.Nauk {\bf 173}, 1382 (2003).
\item{} V.K. Shumny, Fundamental biology and new technology,
             SD RAS Scientific Session, 2002, Frontiers of Science,
             To 50th Anniversary of the DNA double helix discovery,
             SD RAS, Novosibirsk, 2003.
\item{} E.I. Kolchinsky, Neocatastrophism and selectionism: the eternal dilemma
             or a possibility of synthesis ? SPb, Nauka, 2002;
        Yu.V. Natochin, review, RAS Bulletin, 2003, Vol. 73, \#6, p.555.
\item{} A. Poincar\'e, {\it On Science}, Moscow, Nauka, 1983,
           A collection of publications and reports about the philosophy of
           Science, in Russian.
\item{} M.N. Rutkevich, In search for optimal strategy.
             To congress "RIO + 10" in Johannesburg,
             RAS Bulletin, 2002, Vol. 72, \#10, p.934.

\end{enumerate}
\end{document}